\documentstyle[12pt]{article}
\begin{document}

\begin{center}

{\Large

{\bf  Nonlinear acousto-electric transport in a two-dimensional
electron system  }}

\vskip 0.6cm

A.~O.~Govorov$^*$ and A.~V.~Kalameitsev   \\
{\it Institute of Semiconductor Physics,
Russian Academy of Sciences, \\
Siberian Branch,
630090 Novosibirsk, Russia}
\vskip 0.5cm

M.~Rotter, A.~Wixforth, and J.~P.~Kotthaus
\\
{\it Sektion Physik der Ludwig-Maximilians-Universit\"at
and Center for NanoScience,
Geschwister-Scholl-Platz 1,
D-80539 M\"{u}nchen, Germany }

\vskip 0.5cm

K.-H.~Hoffmann  and N.~Botkin  \\
{\it CAESAR, Friedensplatz 16, D-53111 Bonn, Germany}

\end{center}

\vskip 1.5 cm
We study both theoretically and experimentally the nonlinear interaction
between an
intense surface acoustic wave and a two-dimensional electron plasma in
semiconductor-piezocrystal hybrid structures.  The experiments on
hybrid systems exhibit
strongly nonlinear acousto-electric effects. The plasma turns into moving electron
stripes, the acousto-electric current reaches its maximum, and the sound absorption
strongly decreases. To describe the nonlinear phenomena, we develop a
coupled-amplitude method for a two-dimensional system in the strongly nonlinear
regime of interaction. At low electron densities the absorption coefficient decreases with
increasing sound intensity, whereas at high electron density the absorption coefficient
is not a monotonous function of the sound intensity.
High-harmonic generation coefficients as a function of the sound intensity have
a nontrivial behavior.
Theory and experiment are found to be in a good agreement.

\vfill
* E-mail: govor@isp.nsc.ru

\newpage

\section*{Introduction} 

The interaction between surface acoustic waves (SAW's)
and mobile carriers in quantum wells is an important method
to study the dynamic properties of two-dimensional (2D) systems.
The SAW can trap carriers and induce acoustic charge
transport as has been investigated
in a number of systems in view of possible device applications  \cite{ACT}.
Also, the SAW-method was applied to study the quantum
Hall effects \cite{Achim89,Willet,Drichko},
electron transport through a
quantum-point contact \cite{Talyanskii},
lateral nanostructures \cite{Nash},
and commensurability effects
in a 2D system  \cite{Shilton95}. However, all
those experiments have been done in the regime of
small signals and  linear interaction.
A recent Letter by {\it Rotter et al.} \cite{Rotter99}
reports strongly nonlinear acousto-electric effects
in a 2D electron gas (2DEG), which become possible in hybrid
structures based on $A_3B_5$ semiconductors and $LiNbO_3$
[9-11].        
In these experiments
an intense SAW breaks  a 2DEG into moving electron stripes and
all characteristics of the acousto-electric interaction are strongly modified
as compared to the linear case. In modern hybrid structures  \cite{Rotter99APL}
the SAW-induced potential amplitude can
become comparable with the band-gap of a semiconductor.
The previous Letter \cite{Rotter99} on nonlinear effects in
the hybrid structures with a 2DEG includes  a brief qualitative analysis.
Here we present  a detailed theoretical study of nonlinear
acousto-electric effects in a 2D electron system  and
develop  a coupled-amplitude method for intense SAW's
interacting with a 2DEG.  Using our theoretical results we can
explain  main experimental observations.
Including the effect of electron diffusion
we find a good quantitative agreement between theory and experiment
for the case of the SAW absorption coefficient in the nonlinear   regime.

Nonlinear acoustic waves in bulk piezocrystals
with free carriers were discussed in a number of theoretical
papers [12-16].       
The application
of a dc voltage to the crystal can result in the current
amplification of sound \cite{Gurevich,Tien,White}
and in the formation of stationary nonlinear waves
\cite{Gurevich,Tien,Gulyaev1}. Analytic  results in the theory of
nonlinear acoustic waves in
bulk piezocrystals with free carriers
were mostly obtained in the limit of
small amplitudes or for the case of very intense acoustic waves
\cite{Gurevich,Beale,Gulyaev1}. For nonlinear SAW's in
crystals with a 3D electron gas, a theory was developed in the
limit of very high amplitudes, when the SAW bunches
electrons near the crystal surface \cite{Gulyaev2}.
Another theoretical aspect related to the generation of
the second harmonic of a SAW was studied in Ref. \cite{2omega} by using
the coupled-amplitude method and perturbation  theory. A theory of
acousto-electric interactions in a 2D electron system was developed
mostly for the linear regime of interaction \cite{lineartheory,Chaplik}.

Here, we study both theoretically and experimentally
the transition from the linear regime of
the acousto-electric interaction to the limit of strongly
nonlinear effects in a 2DEG.
Our theoretical results are applied for a description
of experimental data on hybrid structures  \cite{Rotter99,Rotter99APL}.
We pay attention to  density-dependences of the absorption coefficient and the SAW-velocity shift because the electron density is a tunable parameter in experiments on 2D systems.
Such dependences were not discussed in detail in the context of 3D systems
[12-16].        
It turns out that for low densities, the absorption coefficient  is a decreasing function of the sound intensity caused by the trapping of electrons in the SAW piezoelectric potential.
At sufficiently high electron density, however, the absorption coefficient
is a non-monotonous function of the sound intensity.  This behavior can be understood in terms of a dynamical screening effect. Also, our quantitative analysis shows that the absorption coefficient at  room temperature is strongly reduced due to electron diffusion. A non-monotonous behavior with increasing sound intensity was also found for the intensities of higher harmonics in a short device.

The coupled-amplitude method with the introduction of  fast and slow
variables was used before for a description of bulk systems \cite{Tien,Butcher}.
However, the formulas from the bulk theory can not be directly applied to surface waves because of the
complex character of the lattice vibrations in a SAW. To develop the coupled-amplitude method  in a 2D system, we introduce a local velocity shift and a local absorption coefficient in an integral form through an electron current, a SAW potential, and an electro-mechanical coupling coefficient of a microstructure. The resulting formulas can then be applied to any type of SAW's interacting with a 2DEG. Moreover,  using our approach, we can find
solutions for higher harmonics  for arbitrary SAW intensity. This is in contrast with perturbation methods developed earlier for 3D systems and valid at small SAW intensities \cite{2omega}. At very high intensities,  we find  analytical asymptotic dependences  for  the high-harmonic intensities in a 2D system.

As has been shown before, the linear approximation
holds when $\delta n \ll N_s$, where
$N_s$ is the equilibrium 2DEG density and
$\delta n$ is a density perturbation due to a SAW.
In the linear theory, the absorption coefficient $\Gamma^0$ and
the SAW-velocity shift due to the 2DEG $\delta v_{s}^0$
are given by the well-known relation
\cite{lineartheory,Chaplik}

\begin{eqnarray}
\label{Eq1}
-\frac{\delta v_{s}^0}{v_s^0}+i\frac{\Gamma^0}{2q}=
\frac{K^2_{eff}(q)}{2}\frac{i\sigma_0/\sigma_m}{1+i(\sigma_0/\sigma_m+D_eq/v_s^0)},
\end{eqnarray}
where $q$ is the SAW wave vector,
$v_s^0$ is the sound velocity in the absence of a 2DEG,
$\sigma_0$ and $D_e$  are the 2DEG conductivity and the diffusion coefficient,
respectively.  $ K^2_{eff}$ is the effective electro-mechanical coupling coefficient.
$\sigma_m(q)=v_{s}^0\epsilon_{eff}(q)/(2\pi)$, where $\epsilon_{eff}(q)$ is the effective
dielectric constant in a 2D system. In most piezoelectric crystals, the coupling
$K^2_{eff}\ll1$. In $GaAs$, $K^2_{eff}=0.00064$,
whereas in the hybrid structures as studied here,
it is two orders of magnitude larger, in the range of $0.01-0.05$ \cite{Rotter98},
but still much less then unity.
The goal of this paper is to describe  the acousto-electric interaction for the case of
large amplitude SAW's, when $\delta n \sim N_s$ and the perturbation theory is not
longer valid. At the same time, the coupling $K^2_{eff}$ will be assumed to be much
less than unity. Below, we will  generalize the results following from
Eq.~\ref{Eq1} for the strongly nonlinear case, when $ \delta  n\sim N_s$.

The paper is organized in the following way. In the first section, we will give
the general equations for SAW's on a piezoelectric crystal. The second section is devoted to a
coupled-amplitude method developed for the case of intense SAW's.
The third and fourth sections are about  phenomena related to
large-amplitude SAW's in a 2DEG. Then, we will discuss experimental data on
the hybrid structures and apply our theoretical results for the
interpretation of experiments.

\section{ Model and general equations}

In usual GaAs-based microstructures it is very difficult to realize SAW's with
high-amplitude potentials  because the electro-mechanical
coupling in GaAs is relatively weak.  A strong
piezoelectric interaction can be achieved in hybrid structures
\cite{Rotter99,Rotter97} (Fig.~1). Those consist of
a semiconductor layer being bonded to
a piezoelectric host crystal, in our case, $LiNbO_3$ \cite{Rotter99,Rotter97}.
The semiconductor layer contains an $InGaAs-AlGaAs$
quantum well (QW) with a high-quality 2DEG, to which Ohmic
contacts are formed. The distance between the QW and the piezocrystal is
only  $32~nm$,
whereas  the distance between the QW and the top transport gate is $d=450~nm$.
In our model $x$ and $y$ are the in-plane coordinates and $z$ is the normal one.
The QW plane corresponds to $z=d$, and
the SAW travels in the $x$-direction (see Fig.~1). By changing
the transport-gate voltage  $V_t$  one can tune the electron density $N_s(V_t)$
in a QW.  In this structure,
traveling SAW's can induce very strong piezoelectric fields in the
semiconductor layer
due to the strong piezoelectricity of the host $LiNbO_3$   crystal.  A SAW is
induced and detected by the metallic interdigital transducers,
$IDT1$ and $IDT2$, respectively, at room temperature  \cite{Rotter99}.
The acousto-electric current is measured between two Ohmic contacts labeled
$1$ and $2$ in Fig.~1.

SAW's in a piezoelectric crystal with a 2D plasma are described by the
system of equations \cite{Auld}

\begin{eqnarray}
\label{u1}
\rho \ddot{u}_i=
c_{iklm}\partial_{m}\partial_{k}u_l+p_{lik}\partial_{l}\partial_{k}\phi
\\
\label{u2}
\epsilon(x_3)\partial_i\partial_i\phi-4\pi p_{ikl}\partial_i\partial_lu_k=
-4\pi en\delta(x_3-d),
\end{eqnarray}
where $e=-|e|$ is the electron charge,
$\rho$ is the mass density, and
$\epsilon$ is the dielectric constant.
Further, $c_{iklm}$ is the elastic tensor,
$p_{lik}$ is the piezoelectric tensor,
$u_k(x,z,t)$ is the lattice displacement, and
$\phi(x,z,t)$ and $n(x,t)$ are the electrostatic potential and
the 2D electron density, respectively.
In Eqs.~\ref{u1},\ref{u2}, we have used the notations
$\partial_lf=\partial f/\partial x_l$ and $\dot{f}=\partial f/\partial t$,
and the sum convention for repeated indexes.
$t$ is the time, and  $x_3=z$, $x_2=y$,
and $x_1=x$ are the coordinates.

In our geometry, the SAW propagating in the x-direction
is a purely Rayleigh wave, in which only two components of
the displacement, $u_x$ and $u_z$, are nonzero.
Hence, the electric field ${\bf E}$ is also polarized in the
$(xy)$-plane.  This case corresponds to the hybrid structures
studied in experiments \cite{Rotter99,Rotter97,Rotter98},
where the $128^0$-rotated $Y$-cut
of $LiNbO_3$ is used. The surface of the thin $GaAs$ film is
$(001)$. The SAW propagates in the $[110]$-direction of
$GaAs$ and $X$-direction of $LiNbO_3$.

The electron 2D plasma is described by the usual hydrodynamic
equations

\begin{eqnarray}
\label{n1}
e\dot{n}(x,t)+\frac{\partial j(x,t)}{\partial x}=0,
\\
\label{js}
j=-\sigma\frac{\partial\phi(x,d,t)}{\partial x}-
eD_e\frac{\partial n(x,t)}{\partial x},
\end{eqnarray}
where $j$ is the 2D electron current,
$\sigma=|e|\mu n_s(x,t)$ is the 2DEG conductivity,
$\mu$ is the mobility,
and $D_e$ is the electron diffusion coefficient.
Eq.~\ref{js} is valid in the long wavelength limit,
when $ql_e\ll1$, where $l_e$ is the
electron mean free path.

The wave equations (\ref{u1},\ref{u2})  should be solved together with the
boundary conditions at the surface $z=0$ and at the interface $z=d$.
At $z=0$, these conditions are the following: $\phi(z=0)=0$ and
$\sigma_{zi}=c_{zilm}(\partial_mu_l+\partial_lu_m)/2+p_{lzk}\partial_l\phi=0$.
Here $\sigma_{zi}$ is the $z$-component of the stress. The top metal gate
is thin and does not influence the boundary condition for
the stress tensor. At the semiconductor-piezocrystal interface,
$\phi$, $u_i$, and  $\sigma_{zi}$ should be  continuous,
and $D_z(z=d+\delta)-D_z(z=d-\delta)=4\pi en$, where
$\delta\rightarrow0$ and
$D_i=-\epsilon(z)\partial_i\phi+2\pi p_{ikl}(\partial_k u_l+\partial_l u_k)$
is the electric displacement. For simplicity, we assumed above that
the 2DEG is located directly at the semiconductor-piezocrystal interface.

\section{ Coupled-amplitude  method}

The system of nonlinear equations (\ref{u1}-\ref{js})
can be simplified in the limit of weak electro-mechanical coupling,
$K^2_{eff}\sim  p^2/(c\epsilon)\ll 1$. In this limit, we can introduce
two coordinates, the "slow" variable $x$ and the "fast" variable
$x_1=x-v_s^0t$ \cite{Tien,Butcher}.
The solution is of the form (see Appendixes~1 and 2),

\begin{eqnarray}
\label{u}
{\bf u}(x,z,t)={\bf u}(x,x_1,z)={\bf u}_0(x,z)+
\sum_{n=1,2,...}
a_n(x) {\bf U}^{0}[z;q_n+\delta q_n(x)]e^{iq_nx_1}+c.c.
\\
\label{phi}
\phi(x,z,t)=\phi(x,x_1,z)=\phi_0(x,z)+\sum_{n=1,2,..}
a_n(x)\Phi^0[z;q_n+\delta q_n(x)]e^{iq_nx_1}+c.c. ,
\end{eqnarray}
This solution is written as a sum of
harmonics with wave vectors $q_n=nq$, where $q>0$ is the
wave vector of the initially generated SAW near $x=0$.
The vector
${\bf A}^0[z;q]=\Bigl( {\bf U}^0[z;q]; \Phi^0[z;q] \Bigr)$
and the quantity $\delta q_n$
are determined by a linear system of equations as given in
the Appendix~2.  The envelope functions $a_n(x)$
are slowly changing on
the scale of $\lambda=2\pi/q$.
It is assumed that the SAW intensity related to
the vector ${\bf A}^0[z;q]$ is unity and thus
the total SAW intensity  is $I_{saw}=\sum_{n=1,2,..}|a_n(x)|^2$.
The functions ${\bf u}_0(x,z)$ and $\phi_0(x,z)$ describe
the static spatial distributions, that can be induced by a SAW.

The functions $j(x,x_1,t)$ and $n(x,x_1,t)$
can be written in the standard form
$f(x,x_1,t)= f_0(x)+\sum_{n=1,2,...} f_n(x)e^{iq_nx_1}+c.c. $,
where $f_n(x)$ is an envelope function.
Also, the electric field ${\bf E}$ can be written in the way similar to
Eq.~\ref{phi}.

The solution (\ref{u},\ref{phi}) is a sum of linear-like
SAW's, which slowly vary in the amplitude and in the $z$-profile.
The $z$-distribution of lattice displacement ${\bf U}^0$
is a sum of exponential functions
$\exp{[-\gamma_j(q_n+\delta q_n(x))z]}$, where $\gamma_j(q)$ are
the coefficients depending also  on material constants \cite{Auld}.
On short distances ($\sim\lambda$)
the envelope functions $a_n(x)$
can be regarded as constants and we can solve
the equations (\ref{u1}-\ref{js})  considering only the "fast"
variable $x_1$.
$n_n(x)$ and $\phi_n(x)$ should be considered here as
the functions of the parameters $a_1, a_2,...$, $n_0$, and $E_0$.
Then, having $n_n$ and $\phi_n$ as functions of
$a_1, a_2,...$, $n_0$, and $E_0$,
we can find the behavior of $a_n(x)$ on long-range scale,
$x\sim1/\Gamma^0\gg\lambda$.

The electrostatic potential is written in a self-consistent way:

\begin{eqnarray}
\label{self}
\phi(x,x_1,z)=\phi^{ind}+\phi^{saw},
\end{eqnarray}
where $\phi^{ind}$ and $\phi^{saw}$
are the potentials induced by a 2DEG and by
piezoelectric charges of a SAW.
Using Eq.\ref{u2} we write

\begin{eqnarray}
\label{sep1}
\epsilon(z)\partial_i\partial_i\phi^{ind}=-4\pi en\delta(z-d)
\\
\epsilon(z)\partial_i\partial_i\phi^{saw}=4\pi p_{ikl}\partial_i\partial_lu_k.
\label{sep2}
\end{eqnarray}
$\phi^{ind}$ and $\phi^{saw}$ can be expressed by the
harmonic amplitudes $\phi^{ind}_n$ and $\phi^{saw}_n$.
For example,
$\phi^{ind}=
\phi^{ind}_0(x,z)+\sum_{n=1,2,...}\phi^{ind}_n(x,z)e^{iq_nx_1}+c.c.$.
In the limit
$\Gamma_{max}\lambda\sim K_{eff}^2\ll 1$,
we find from Poisson's equation (see Appendix~1)  and
from the conservation of charge

\begin{eqnarray}
\label{j}
\phi^{ind}_n(x,d)=\frac{2\pi e n_n(x)}{\epsilon_{eff}(q_n)q_n},
\hskip0.5 cm
E^{ind}_{nx}(x,d)=-iq_n\phi^{ind}_n(x,d),
\\   \nonumber
j_n(x)= v_s^0en_n(x),
\end{eqnarray}
where $n=1,2,3...$ .
Here $\epsilon_{eff}(q)= [\epsilon_p+\epsilon_s coth(|q|d)]/2$ is
the effective dielectric constant including the gate electrode effect,
and $\epsilon_p$ and $\epsilon_s$ are the dielectric constants of a
host piezocrystal and a semiconductor, respectively \cite{ChaplikPL}.

The $n$th-harmonic of the SAW-potential $\phi^{saw}_n$
is given only by  $a_n(x)$ and by material constants and
can be easily found from the Poisson equation (\ref{sep2}).
At $z=d$ we have $\phi^{saw}_n(x,d)=C_na_n(x)$, where
the coefficient $C_n$ depends on the geometry and
the material constants. For example,
in a crystal of the type of $GaAs$,  $C_n=p_4g(q_n)$, where
$g(q_n)$ is a complicated function of $q_n$.  Below, we will
give the necessary relations for the hybrid structures.

To find the harmonic amplitudes $n_n$ and $\phi_n$, we have to solve
Eqs.~\ref{u2}, \ref{n1}, and \ref{js}
regarding the "slow" variable $x$ as the constant.
The slowly-varying quantities $n_n(x)$ and $j_n(x)$
can be found as Fourier components of the solution $n(x_1,x)$
from  a "fast" equation in terms of $x_1$.
In a self-consistent approach, the electron density $n(x_1,x)$ is
determined by the SAW-induced potential at $z=d$,
$\phi^{saw}(x_1,x,d)=
\phi^{saw}_0(x,d)+ \sum_{n=1,2,...} C_n a_n(x)e^{iq_nx_1}+c.c.$,
through a non-linear equation \cite{Rotter99}.
By using Poisson's equation, the results of Appendix~1,
and Eqs.~\ref{n1} and \ref{js}, we obtain

\begin{eqnarray}
\label{int}
|e|n(x_1,x)\mu[
-\frac{d}{dx_1}\{\int^{+\infty}_{-\infty}dx'_1 G(x_1-x'_1) n(x'_1,x) \}+
E^{saw}(x_1,x)]
\\
\nonumber
-eD_e\frac{dn(x_1,x)}{dx_1}-ev_{s0} n(x_1,x)=b_0 ,
\end{eqnarray}
\begin{eqnarray}
\label{int2}
G(x_1-x'_1)=
e\int^{+\infty}_{-\infty}dk\frac{e^{-ik(x_1-x'_1)}}{|k|\epsilon_{eff}(|k|)},
\end{eqnarray}
where $b_0$ is a constant, which occurs after one integration in the
conservation-of-charge equation (\ref{n1}).  Taking into account
$\Gamma^0\lambda\ll1$,  we assume that
the solution of Eq.~\ref{int} is  periodic
in $x_1$, $n_s(x_1,x)= n_s(x_1+\lambda,x)$. Also,
$<n_s(x_1,x)>= \int_0^{\lambda}n_s(x_1,x)dx_1/\lambda=n_0(x)$, where
$n_0(x)$ plays the role of a "local" 2D density.

The functions $a_n(x)$ are
connected by the system of nonlinear equations
(see Appendix 2)

\begin{eqnarray}
\label{a}
\frac{da_n}{dx}=i\delta q_n[a_1,a_2,...;n_0,E_0] a_n .
\end{eqnarray}
where

\begin{eqnarray}
\label{dq}
\delta q_n[a_1,a_2,... ;n_0,E_0]=-\frac{K^2_{eff} (q_n)}{2}
\frac{2\pi}{\epsilon_{eff}(q_n)} \frac{en_n(x)}{\phi_{n}^{saw}(x)}=
i\frac{K^2_{eff} (q_n)}{2q_n\sigma_m(q_n)}
\frac{j_n(x)E_n^{saw*}}{|\phi_{n}^{saw}(x)|^2}
\end{eqnarray}
and $n=1,2,3...$ .

The local velocity change of a SAW $\delta v_n$ and the local
absorption coefficient $\Gamma_n$ can be
expressed by $\delta q_n $:
\begin{eqnarray}
\label{dq1}
\delta q_n =-q_n\frac{\delta v_n}{v_s^0}+i\frac{\Gamma_n }{2}.
\end{eqnarray}
Using the relation

\begin{eqnarray}
\label{Keff}
K_{eff}^2(q_n)= \frac{2|E_n^{saw}|^2\sigma_m(q_n)}{q_n I_n},
\end{eqnarray}
(see Ref.~\cite{Keff}) and Eqs.~\ref{j} we write

\begin{eqnarray}
\label{G-v}
\frac{\delta v_n(x)}{v_s^0}=
\frac{<\tilde{\phi }^{saw}_n (x_1,x)j(x_1,x)>}{2 I_n(x)} \\
\label{G-v2}
\Gamma_n(x)=\frac{<\tilde{E}^{saw}_n(x_1,x)j(x_1,x)>}{I_n(x)}.
\end{eqnarray}
Here we use the notations $<f(x_1,x)>= \int_0^{\lambda}f(x_1,x)dx_1/\lambda$, \
and \ $\tilde{f}_n(x_1,x)= f_n(x)exp(iq_nx_1)+c.c.$ .

In this section, we have assumed that $K^2_{eff}\ll1$ and
neglected the terms $d^2a_n/dx^2\propto\delta q_n da_n/dx\propto K^4_{eff}$
and $d\delta q_n/dx\propto K^4_{eff}$.
The static electric fields $\phi_0(x,z)$ in Eq.~\ref{phi},
which can be induced by a SAW,
will not play an important role in this paper because we will consider the case with
no voltage applied to the Ohmic contacts ($V_1=V_2$) and a relatively short device
with $L\ll1/\Gamma^0$.
Thus, $\phi_0(x,d)\simeq const$ and
the static electric fields $E_{0x}(x,d)$ can be neglected.
Eqs. \ref{a} and \ref{dq}  can now be applied to various
types of SAW's by introducing specific electro-mechanical
coupling coefficients.

\section{Acousto-electric transport in a two-dimensional plasma}

Here, we intend to consider the SAW absorption and the acoustic charge transport
in the regime $V_1=V_2$.
The behavior of a 2DEG in an intense SAW at small distances ($\sim\lambda$)
can be assumed to be periodic and is described by nonlinear equation (\ref{int}).
At long distances ($\gg\lambda$) the SAW behavior is
determined by complex amplitudes $a_n(x)$, that can be found by
Eqs.~\ref{a} and \ref{dq}. First, we define the boundary conditions
at $x=0$: $a_1(0)=\sqrt{I_1(0)}$ and $a_n(0)=0$ for $n=2,3,...$.
Here, $I_1(0)$ denotes the SAW intensity
generated by IDT1.  In this case, for a relatively short sample
the SAW contains
mostly the fundamental SAW harmonic $n=1$ with a small admixture of
the higher harmonics $n=2,3,...$.
So, it follows from Eqs.~\ref{a} and \ref{dq}

\begin{eqnarray}
\label{x-Bihav1}
a_1(x)=a_1(0)\Bigl(1 + ix\delta q_1[a_1(0),0,0,...;n_0,0]\Bigr) .
\end{eqnarray}
We assume that $[I_1(0)-I_1(L)]\ll I_1(0)$, where
$L$ is the length of a semiconductor film and
$I_1(L)$ is the intensity detected at the IDT2.
The latter is valid in a short sample where $\Gamma_1L\ll1$.
The first-harmonic absorption coefficient per unit length,
which is measured in the experiments,
is $[I_1(0)-I_1(L)]/(I_1(0)L)=
Im\Bigl(\delta q_n[a_1(0),0,0,...;n_0,0]\Bigr)=\Gamma_1(a_1,0,0,...;n_0,0)\equiv\Gamma_1$
and the velocity shift
is $\delta v_1(a_1,0,0,...;n_0,0)/v^0_s=Re\Bigl(\delta q_1[a_1,0,0,...;n_0,0]/q\Bigr)\equiv
\delta v_1/v^0_s$.

The amplitudes of higher harmonics in a short sample turn out to be
given by:

\begin{eqnarray}
\label{x-Bihav2}
a_n(x)=ix \lim_{a_n\rightarrow0}{a_n\delta q_n[a_1(0),0,0,..,a_n,...;n_0,0]}=
ix\frac{\pi K_{eff}^2(q_n)}{\epsilon_{eff}(q_n)}\frac{en_n(0)}{C_n},
\end{eqnarray}
where $n=2,3,4,... $ . Here, we  assume that $[I_1(0)-I_1(L)]\gg I_n$, which is valid for a short sample where $\Gamma_1L\ll1$,
and we take into account that for the strongly nonlinear case $\Gamma_1\sim\Gamma_n$.
The intensity of the $n$th-harmonic is given by $I_n(x)= |a_n(x)|^2$ and

\begin{eqnarray}
\label{Cn}
|C_n|^2=\frac{|\phi_n^{saw}|^2}{|a_n|^2}=
\frac{K_{eff}^2(q_n)}{2q_n\sigma_m(q_n)}
\end{eqnarray}
(see Eq.~\ref{Keff}).
By using Eq.~\ref{x-Bihav2}
we find the intensities of high harmonics at $x=L$:
$I_n(L)=\lim_{a_n\rightarrow0}{|a_n\delta q_n[a_1,0,0,...,a_n,...;0,0]|^2}L^2=
\pi L^2(K_{eff}^2/\epsilon_{eff})q_nv_s^0e^2|n_n(0)|^2$.

To calculate $\Gamma_1$ and  $\delta v_1$,
we numerically  solve  Eq.~\ref{int} for the parameters $a_1\neq0$ and $a_n=0$ when
$n=2,3,...$. Then we can find the Fourier components of $n(x_1,a_1)$.
We now calculate the electron density $n(x_1,a_1)$ in the
long wave-length limit $qd\ll1$. In this limit  the function
given by Eq.~\ref{int2}  is reduced to
$G(x_1-x_1')=(4\pi ed/\epsilon_s) \delta(x_1-x_1')$ and
Eq.~\ref{int} now writes as

\begin{eqnarray}
\label{int3}
|e|n(x_1,a_1)\mu
\{-\frac{4\pi ed}{\epsilon_s}\frac{dn(x_1,a_1)}{dx_1}+
F_1^{saw}cos(qx_1+\chi_1)\}-eD_e\frac{dn(x_1,a_1)}{dx_1}\\
\nonumber
-ev_s^0n(x_1,a_1)-b_0=0,
\end{eqnarray}
where the SAW-induced piezoelectric field is taken in the form
$E^{saw}=F_1^{saw}cos(qx_1+\chi_1)$, and $\chi_1$ is the phase of the first
SAW harmonic. For simplicity, we put in the following $\chi_1=0$.
The constant $b_0$ in Eq.~\ref{int3} is directly connected with
the kinetic motion of a SAW and thus vanishes for $v_s^0\rightarrow0$.
Dividing Eq.~\ref{int3} by $n$ and then integrating over
$x_1$ we get

\begin{eqnarray}
\label{int4}
|e|\mu\{-\frac{4\pi ed}{\epsilon_s}n(x_1)+\frac{F_1^{saw}}{q}\sin{(qx_1)}\}
-eD_eln[n(x_1)]= \\
const_1+ev_s^0x_1+b_0\int_0^{x_1}\frac{dx'_1}{n(x'_1)}.
\nonumber
\end{eqnarray}
The left-hand side of the above equation is periodic and thus
the right-hand side should be periodic as well.  It implies that
$b_0\int_0^{\lambda}dx'_1/n(x'_1)=-ev_s^0\lambda$.

A numerical solution of Eq.~\ref{int3} $n(x_1)$ at $T=300 K$ and
at various averaged densities $N_s=<n(x_1)>$ is shown in the inset of Fig.~3.
Here, we use the following parameters:
$\lambda=33~\mu m$, $\mu=5000~cm^2/Vs$, $D=\mu (KT/e)$,
$\epsilon_s=12.5$, and $v_s^0=3.8 *10^5~cm/s$.
It is seen that that with decreasing $N_s$ the formerly homogenous
2DEG turns into moving electron stripes.

It follows from Eqs.~\ref{G-v},\ref{G-v2}:

\begin{eqnarray}
\label{abs-vel}
\frac{\delta v_1}{v_s^0}=-\frac{<\Phi_1^{saw}sin(qx_1)j(x_1)>}{2I_1},
\hskip 0.7cm
\Gamma_1=\frac{<F^{saw}_1cos(qx_1)j(x_1)>}{I_1}.
\end{eqnarray}
The calculated absorption coefficient
$\Gamma_1$ as a function of the electron density $N_s$ for various potential
amplitudes
$\Phi_1^{saw}= F_1^{saw}/q$ is shown in Fig.~2.
It was calculated from Eqs.~\ref{abs-vel} and the numerical solution for  $n(x_1)$.
The potential amplitude $\Phi_1^{saw}$ can be easily connected with
the SAW intensity $I_1$ by using Eqs.~\ref{Keff} and
\ref{Cn} with $K_{eff}^2$ found
numerically in Ref.~\cite{Rotter97}.
As an example, in the hybrid structures
at SAW frequencies $f=114 ~ MHz$ and $340  ~ MHz$, $K_{eff}^2$ is
about $0.015$ and $0.035$, respectively.

We see in Fig.~2 that with increasing $\Phi_1^{saw}$ the absorption coefficient
in general decreases and its maximum  is shifted to
the higher values of $N_s$.  This nonlinear behavior can be understood qualitatively as
follows.
For the densities $N_s\ll N_s^{max}$, the electron plasma forms the moving
charge stripes and the electron velocity $j/(eN_s)$ is very close to
its maximum $v_s^0$. Here $N_s^{max}$ denotes the density corresponding to
the maximum of the function $\Gamma_1(N_s)$ at fixed $\Phi_1^{saw}$.
We will show below  that  in the case $N_s<N_s^{max}$
$\Gamma_1\simeq |e|N_s(v_s^{0})^2/(I_1\mu)\propto N_s/I_1$ .
This asymptotic behavior follows from the Weinreich relation \cite{Weinreich}.
In the region $N_s\gg N_s^{max}$, the electric current $\sigma E_x$ decreases
with increasing $N_s$
because of the screening effect. Thus, the absorption
coefficient decreases as well.

In Fig.~3 we plot $\Gamma_1$ as a function of  the SAW potential
$\Phi_1^{saw}$ for a fixed 2DEG density.
For electron densities less than about $10^{10}cm^{-1}$,
the function  $\Gamma_1(\Phi_1^{saw})$ is always decreasing.  At higher
densities $\Gamma_1(\Phi_1^{saw})$ has a maximum. We attribute this behavior
to the screening effect in a 2DEG.     At high density and small $\Phi_1^{saw}$,
the absorption is strongly suppressed because of screening.
The intense SAW, however,  modulates
the 2DEG and, consequently,  reduces screening.
Thus, the absorption coefficient starts to increase with
increasing $\Phi_1^{saw}$, when $\Phi_1^{saw}$ is not so large. At  larger
$\Phi_1^{saw}$, the plasma becomes broken into stripes and $\Gamma_1$ decreases with
$I_1$ as $1/I_1$ \cite{ Rotter99,Gulyaev2}.

The velocity shift $\delta v_1$ numerically calculated by Eqs.~\ref{abs-vel}
is shown in Fig.~4. With increasing SAW intensity   $\delta v_1$ increases
and approaches to the velocity $v_s^0$ that corresponds to the case of a totally depleted
2DEG.  By analyzing Eqs.~\ref{abs-vel} and the function $j(x_1)$,
we can see that, in the limit $I_1\rightarrow\infty$,
$\delta v_1\propto1/\Phi_1^{saw}\propto1/\sqrt{I_1}$, like in a 3D plasma
\cite{Gurevich}.

In Fig.~5 we show results of numerical calculations for the intensities of the high harmonics
$I_n/I_{n,max} = |n_n|^2/N_s^2$,  where

\begin{eqnarray}
\nonumber
I_{n,max}=I_n(\Phi_1^{saw}\rightarrow\infty)=
\pi L^2\frac{K_{eff}^2}{\epsilon_{eff}}q_nv_s^0e^2N_s^2.
\end{eqnarray}
This formula was obtained taking into account that $|n_n|\rightarrow N_s$ when
$\Phi_1^{saw}\rightarrow\infty$.
The quantities $I_n$ were calculated from the Fourier components of the function
$n(x_1)$.
At small amplitudes $\Phi_1^{saw}$, $I_n\propto(\Phi_1^{saw})^n$, being typical
for weak nonlinearity. At larger  $\Phi_1^{saw}$, the behavior of $I_n$ is
quite complex and
strongly differs from the weak-nonlinearity behavior. At very large  $\Phi_1^{saw}$,
the high-harmonics intensities tends to saturate: $I_n\rightarrow I_{n,max}$.
At higher electron densities the saturation of $I_n$ occurs at
larger SAW potentials (see Fig.~5).

\section{ Analytic results for acousto-electric effects in a
2D electron plasma}

In this section, we will give some analytic expressions
describing various acousto-electric effects
in a system with a 2DEG. For simplicity, we will
not take into account  diffusion, hence  assuming $D_e=0$
in the formula for the current (\ref{js}).
This assumption is well justified at large electron densities or at low temperatures.
It is convenient to start with the Weinreich relation
for a 2D system in the nonlinear regime  \cite{Gurevich,Weinreich}.
Assuming that $n(x_1)$ and $j(x_1)$ are periodic functions
we can rewrite Eq.~\ref{int} in the form

\begin{eqnarray}
\label{curr}
j(x_1)-v^0_sen(x_1)=b_0 .
\end{eqnarray}
Obviously, it is valid in the limit $K^2_{eff}\ll1$.  Using  Eq.~\ref{curr} we can write
$<j>=|e|\mu<nE_x>=-\mu<jE_x>/v_s^0$.
From this equation we now get the Weinreich relation:

\begin{eqnarray}
\label{Wein}
\frac{<jE_x>}{<j>}=-\frac{v_s^0}{\mu},
\end{eqnarray}
where  $<jE_x>$  is the dissipation  in a SAW.
To get Eq.~\ref{Wein}, we have neglected the averaged electric fields
$<E_x>=0$ assuming the case $V_1=V_2$.

We now consider Eq.~\ref{int3} with $D_e=0$. At fixed $\Phi_1^{saw}$,
there is a critical density, $N_{crit}$, for the formation of stripes in  a 2DEG.
When $N_s<N_{crit}$ the plasma is split into electron stripes. If $N_s>N_{crit}$,
the plasma is continuous but can be strongly modulated in space.
For the case  $N_s<N_{crit}$, the constant $b_0$ in Eq.~\ref{int3} becomes zero
and thus Eq.~\ref{int3} has formally two solutions:

\begin{eqnarray}
\label{N1}
f_1(x_1)=const_2-\frac{\epsilon_s\Phi^{saw}_1}{4\pi|e|d}\sin{(qx_1)}-
\frac{v_s^0}{\mu}\frac{\epsilon_s}{4\pi|e|d}x;  \hskip 0.7cm
f_2(x_2)=0.
\end{eqnarray}
The solution has to be a periodic continuous combination of these
two functions.  It follows from Eq.~\ref{N1} that
in the limit $\sigma_0\gg\sigma_{m}$
$N_{crit}\simeq\Phi_1^{saw}\epsilon_s/(4\pi|e|d)$,
where $\sigma_0=|e|\mu N_s$.

When $N_s<N_{crit}$ the plasma turns into stripes, which means that electrons
are totally trapped and the local electron velocity in a 2DEG reaches its maximum
$j/(eN_s)=v_s^0$. From the Weinreich relation, we get
$\Gamma_1=<jE_x>/I_1=|e|N_s(v_s^0)^2/(\mu I_1)\propto1/I_1$
\cite{Rotter99,Gulyaev1}.

Eqs.~\ref{int3} and \ref{int4} can be used to find an asymptotic formula for $\Gamma_1$ and
$<j>$ in the large density limit when $\sigma_0\gg\sigma_{m}$.
In the limit $\sigma_0\gg\sigma_{m}$,
$n\simeq N_s-(\epsilon_s\Phi^{saw}_1)/(4\pi|e|d)\sin{(qx_1)}$ (see Eq.~\ref{N1}).
By using the above and Eqs.~\ref{curr} and \ref{Wein}, we have in the limit
$\sigma_0\gg\sigma_{m}$ and in the region $\Phi_1^{saw}<\Phi_{crit}$:

\begin{eqnarray}
\label{Ass1}
\Gamma_1=\Gamma_{max}\frac{4\sigma_m}{\sigma_0}(\frac{\Phi_{crit}}{\Phi_1^{saw}})
^2
\Bigl(1-\sqrt{1-[\frac{\Phi_1^{saw}}{\Phi_{crit}}]^2}\Bigr),
\end{eqnarray}
where
$\Gamma_{max}=q K^2_{eff}/4$
and
$\Phi_{crit}=N_s4\pi|e|d/\epsilon_s$.
This equation is valid when
$(\Phi_{crit}-\Phi_1^{saw})/\Phi_{crit}\gg\sigma_m/\sigma_0$.
Eq.~\ref{Ass1} reproduces the numerical
data for $\Gamma_1(\Phi_1^{saw})$ in Fig.~3 at large densities.
The asymptotic formula  (\ref{Ass1}) was given before in
Ref.~\cite{Vyun} without noting the condition
$\sigma_0\gg\sigma_m$. In the linear regime of interaction
$\Gamma_1$ and $\delta v_1$ are given by the formula (\ref{Eq1}).

In the end of this section we consider an asymptotic behavior for
the high-harmonic intensities $I_n$ in the limit
$I_1\rightarrow\infty$. The electron density $n_s(x_1)$
at high $\Phi^{saw}_1$ can be written in a parabolic approximation.
Then, by calculating
the Fourier components $n_n$, we find that
$I_{n,max}-I_n\propto (\Phi^{saw}_1)^{-2/3}\propto I_1^{-1/3}$,
where $n=2,3,...$.

\section{ Comparison with experimental data}

The experiments  involving SAW's were performed on the hybrid
semiconductor-$LiNbO_3$ structures
fabricated by the epitaxial lift-off (ELO) technique developed by
{\it Yablonovich et al.} \cite{Yablonovich}.
The structures contain a $12-nm$-thick high-quality
$In_{0.2}Ga_{0.8}As$ quantum well (QW) embedded in modulation doped
$Al_{0.2}Ga_{0.8}As$ barriers.
In these structures, the thin semiconductor layered system
including a QW was tightly bound to the
lithium niobate host crystal  by
the van der Waals forces \cite{Rotter99,Rotter97,Rotter98}.
The MBE grown quantum well structure is removed from
its native $GaAs$ substrate by etching an $AlAs$ sacrificial
layer below the active semiconductor system.
The thin ELO film with a thickness of only $500 nm$ is then
transferred onto the host $LiNbO_3$ crystal.
The parameters for this structure were described already in the
beginning of Sec.~1. The geometry of the structure
 is shown in Fig.~1. For further details
related to the fabrication procedure of such quasi-monolithic
structures
we refer the reader to Refs. [8-10].        
The experiments were performed  for two SAW
frequencies $f=340~MHz$ and $f=114~MHz$ at room temperature.

The SAW in our experiments can be strong enough to break up
an initially homogenous 2D plasma into moving stripes.
The transition to the regime of moving electron stripes
was directly observed in the experiments on acoustic charge
 transport (ACT)  in samples with
specially-designed injection and detection dates \cite{Rotter99}.
In these experiments the velocity of the ACT-signal  first increases with
the SAW-intensity and  finally saturates  at the sound velocity. The latter manifests the formation of stripes.
Strongly nonlinear effects are also observed in the attenuation data.
The attenuation of a SAW with $f=114~MHz$ for different
intensities is plotted in the insert of Fig.~2 as a function
of the transport-gate voltage, which determines
the averaged electron density in a 2DEG.
At small SAW-intensities, the electronic sound attenuation $\Gamma^0$
as a function of the conductivity $\sigma_0$ is described by the well-known
linear-theory equation (\ref{Eq1}) and exhibits a maximum.
This linear regime is
realized in our experiments at the smallest SAW intensities of about
$-12~dBm$ (insert of Fig.~2).
It is seen from the insert of Fig.~2 that at high SAW amplitudes
the attenuation is strongly suppressed and its maximum is
shifted to higher gate bias or conductivity, respectively.
The experimental data for a SAW with the frequency $f=340~MHz$ look qualitatively
similar to those for $f=114~MHz$ and were given earlier
 in Ref. \cite{Rotter99}.

The nonlinear regime of interaction is described by
the theory given in Secs.~2, 3, and 4.
To quantitatively compare theory and experiment, we now express
the SAW potential amplitude $\Phi_1^{saw}$ through the  input radio frequency  (RF) power $P$. The SAW intensity can be written as
$I_1=I_{saw}=2(P/w)10^{-IL/10}$,
where the width of the transducer $w=0.55~mm$.
The insertion  losses $(IL)$ in the transducers
were measured to be $15~dB$.  Then, the SAW potential
can  be written using Eq.~\ref{Cn} as
$\Phi_1^{saw}=K_{eff}\sqrt{2I_1/(q\sigma_m)}$,
where $K^2_{eff}=0.015$ for $f=114 MHz$ \cite{Rotter98}.
In Fig.~3 we also show the experimentally measured absorption coefficient
at the gate voltage $V_t=-7.5~V$. This voltage corresponds to the maximal
attenuation for the smallest RF power, $-12~dBm$ (see insert of Fig.~2).
From the linear theory we find that the absorption coefficient  is maximal
at $N_s=0.9*10^{10}~1/cm^2$.  One can see from Fig.~3 that
the experimentally measured function
$\Gamma_1(\Phi_1^{saw})$  for $ V_t=-7.5~V $ is in a very good agreement
with the calculated one for $N_s=0.8*10^{10}~cm^{-2}$. Here
we did not use any fitting parameters.  This quantitative agreement becomes  possible
if we account  for the diffusion coefficient. The maximal absorption coefficient  as calculated from the linear theory is $\Gamma^{diff}_{max}=7.6~cm^{-1}$.
Without diffusion this value is about  $\Gamma_{max}=14.3~cm^{-1}$.  Thus, the diffusion strongly suppresses the SAW absorption.

At fixed SAW power and a sufficiently small density $N_s$, the 2DEG is
divided into stripes and $\Gamma_1$ increases with increasing the gate voltage.
In our theory, $\Gamma_1\propto N_s$ in the regime of stripes, which
explains the increase of the attenuation at small gate voltages in the insert of
Fig.~2.
At a sufficiently large gate voltage, the absorption coefficient
as a function of $V_t$ starts to
decrease because of the screening effect in a high-density 2DEG modulated by a SAW.
The interplay of these two effects leads to the shift of the maximum of the
function  $\Gamma_1(V_t)$  shown in  the insert of Fig.~2.

The change of the SAW-velocity due to the electron plasma
is shown in the  insert of Fig.~4. With increasing RF power
the curves in Fig.~4 are again shifted towards larger electron
conductivity  which can be understood in terms of screening.
With increasing  SAW intensity the electron plasma is
strongly modulated or even split into stripes and  the screening of
piezoelectric fields by electrons becomes not so effective.
Thus, the shift of the SAW-velocity due to electrons
decreases with increasing the SAW intensity.
The experimentally  observed shift is in qualitative agreement with
our modeling   shown in Fig.~4. However, our theory does not reproduce the character of
$\delta v_1(V_t)$ in the region of small electron densities.
Likely, the measurement of the SAW velocity is not so sensitive to
a low-density  electron system in comparison
with the attenuation method.

In Fig.~6, we show the quantity  $\Gamma_1P/ I_{ae}$ as a function
of the RF power to verify the Weinreich relation.
The acousto-electric current $<j>=I_{ae}(P)$  was measured in a "short circuit geometry", where the Ohmic contacts are directly (without resistor) connected to the current measurement instrument.  For more details on acousto-electric current measurements we refer to Ref.~\cite{Rotter99APL}.
We see from Fig.~6 that
the ratio $\Gamma_1P/ I_{ae}$ has a weak power-dependence.
Thus, our experimental data are well described by the Weinreich relation.
Slight deviations from the Weinreich relation seen in Fig.~6
can come from the density-dependence of the mobility $\mu(N_s)$, that is
expected to be relatively weak at room temperature. The reason is that
the main electronic scattering mechanism at high temperatures is due to   acoustic phonons and is relatively insensitive to the 2D density.

\section{Conclusions}

The theoretical results obtained in Sec.~2, can also be applied to study
dynamics of SAW's at large distances in a long sample, where
the contribution of high harmonics can be very important
\cite{Balakirev1,Balakirev2}. This long-distance transformation to high
harmonics was studied experimentally for
SAW's interacting with a 3D electron gas of a semiconductor on a
piezocrystal \cite{Balakirev1,Balakirev2}. In the presence of a dc voltage applied
to the crystal, it is possibly to expect the appearance of nonlinear
waves with a stationary profile or with a stationary energy flow
\cite{Gurevich,Tien,Balakirev1}.
In a wave with a stationary energy flow the wave shape is periodically changed
in space \cite{Balakirev1}.  Another scenario can relate to chaotic dynamics in
an acousto-electric system  \cite{Chaos}.
Eq.~\ref{a} can be used to numerically model these phenomena in 2D electron systems
at long distances.

To conclude, we have studied strongly nonlinear acousto-electric
phenomena caused by the interaction between a SAW and
a two-dimensional electron system.
In the experimental  measurements  performed on hybrid
semiconductor-piezocrystal
structures  the SAW attenuation, the SAW velocity  change,
and the acousto-electric current  are strongly modified   in
the nonlinear    regime due to the formation of moving
electron stripes.
By using a coupled-amplitude method we have modeled the decay
and transformation of SAW's in the nonlinear regime.
Using our theoretical results, we could explain
our experimental findings and distinguished between different regimes of
the nonlinear acousto-electric interaction at large SAW intensities.

\section*{ Acknowledgements }
We would like to thank D. Bernklau and H. Riechert for the
fabrication of the excellent MBE material, W. Ruile
for the strong support from
the SAW device side, and A. V. Chaplik and M. K. Balakirev for helpful
discussions.
We gratefully acknowledge financial support by the Volkswagen-Stiftung
and by the Russian Foundation for Basic Research
(grants 99-02-17019 and 99-02-17127).

\section*{ Appendix 1 }

Here we intend to briefly discuss the electrostatics of the hybrid structure.
The spacing between the 2DEG and the top metal gate in the fabricated
structures is much larger than the distance  from the 2DEG
to the $AlGaAs-LiNbO_3$ interface.
Thus, to model the screening effects, we will assume that the 2DEG is
located right on the $AlGaAs-LiNbO_3$ interface. It is convenient to
solve this problem by a Fourier transform in terms of $x$ and by remaining
the vertical coordinate $z$.
The relation between the Fourier components of the electrostatic potential
$W^{ind}[z;k]$  induced by 2D electrons
and the 2DEG density $n_k$ is found from the Poisson equation and from the corresponding boundary conditions,

\begin{eqnarray}
\label{App1}
W^{ind}[z;k]= \frac{2\pi e n_k}{|k| \epsilon_{eff}(k)}G(z),
\end{eqnarray}
where
$\epsilon_{eff}(k)=(\epsilon_p+\epsilon_scoth(|k|d))/2$.
The function $G(z)=\frac{\sinh{|k|z}}{\sinh{|k|d}}$,
when  $0<z<d$, and $G(z)=e^{-|k|(z-d)}$ for $d<z$ \cite{ChaplikPL}.

Taking the electron density in the form of
$n_s(x,x_1,t)=n_0(x)+ \sum_{n=1,2,...} n_n(x)e^{iq_nx_1}+c.c.$,
we can write for the induced electrostatic potential:
$\phi^{ind}(x,x_1,z,t) = \phi^{ind}_0(x,z) + \sum_{n=1,2,...}
\phi^{ind}_n(x,z) e^{iq_nx_1}+c.c.$,
where $q_n=nq$.
Using  Eq.~\ref{App1}, we find
$\phi^{ind}_n(x,z)=2\pi e n_n(x)G(z)/\Bigl(q_n\epsilon_{eff}(q_n)\Bigr)+\delta\phi_n(x,z)$,
where $n=1,2,...$.
The correction
$\delta\phi_n(x,z)\sim da_n(x)/dx\sim\delta q_n\sim K^2_{eff}$
(see Eq.~\ref{dq} for $\delta q_n$) and is small
compared to $W^{ind}[z;q_n]$.  Thus, regarding $n_n(x)$ as the constants, we
obtain at $z=d$: $\phi^{ind}_n(x,d)\simeq2\pi e n_n(x)/[q_n\epsilon_{eff}(q_n)]$ and
$E_{nx}(x,d)\simeq-iq_n\phi^{ind}_n(x,d)$.
In other words, we consider the envelope functions
$n_n(x)$ as constants and solve Poisson's equation in terms of the "fast" variable
$x_1$. Again, it is valid in the limit $\Gamma^0/q\sim K^2_{eff}\ll1$.

As an example, we now calculate the correction  $\delta\phi_n$ in the linear
regime of interaction, when $n(x,x_1,t)=N_s+n_1(x)e^{iqx_1}+n_1^*(x)e^{-iqx_1}$ with
$n_1(x)=\tilde{n}_1 e^{-\Gamma^0x/2-iq\delta v_s^0/v_{s}^0x}$.
From the Poisson equation we find
$\delta\phi_1(x,z) \sim (\Gamma^0/2+iq\delta v_s/v^0_{s})
n_1(x) F(z)$,
where $F(z)\sim1$ for $z\sim1/q$.

\section*{ Appendix 2 }

In order to solve a system of nonlinear equations (\ref{u1}-\ref{js})
we will use some of results from a linear-response theory
\cite{lineartheory,Chaplik}. In a linear theory the total electrostatic
potential and the 2D density can be written as
$W[x,z,t]=W[z,k]e^{ikx-i\omega t}$ and $n(x,t)=n_ke^{ikx-i\omega t}$, respectively.
Here $\omega$ is the SAW frequency.
It is convenient to introduce a quantity
$\Pi_k$ by means of the relation  $en_k=-\Pi_kW[d;k]$, where
$W[d,k]$ is the Fourier component of the electrostatic
potential at $z=d$. Eqs.~\ref{u1},\ref{u2} are now written as

\begin{eqnarray}
\label{App21}
\omega^2\rho u_i+c_{iklm}\partial_{m}\partial_{k}u_l+
p_{lik}\partial_{l}\partial_{k}W[x,z,t]=0,
\\
\label{App22}
-4\pi p_{ikl}\partial_{i}\partial_{l}u_k+
\Bigl(\epsilon\partial_{i}\partial_{i}-4\pi \Pi_k\delta(x_3-d)\Bigr)W[x,z,t]=0.
\end{eqnarray}
Above equations should be solved together with the necessary boundary conditions
considered in Sec.~1.
Then, we rewrite Eqs.~\ref{App21},\ref{App22} in the form

\begin{eqnarray}
\label{App23}
\hat{L}_{lin} {\bf A}_{lin}=0,
\end{eqnarray}
Here $\hat{L}_{lin}$ is a linear operator and
${\bf A}_{lin}=\Bigl({\bf u}[x,z,t],W[x,z,t]\Bigr)=
\Bigl({\bf U}^0[z;k],W^0[z;k]\Bigr)e^{ikx-i\omega t}$.
It follows from the boundary conditions that
$k=q+\delta q_{lin}$ \cite{Chaplik},
where $q=\omega/v_s^0$ and

\begin{eqnarray}
\delta q_{lin}=\frac{qK_{eff}^2}{2}\frac{\Pi_{q}/\Pi^0_q}{1+\Pi_{q}/\Pi^0_q},
\hskip0.5cm
\Pi^0_q= \frac{q\epsilon_{eff}(q)}{2\pi} .
\label{App24}
\end{eqnarray}
Now a solution of Eq.~\ref{App23} can be written as ${\bf A}_{lin}=
{\bf A}^0[z;q+\delta q_{lin}]f_0(x)e^{iqx_1}$,
where $f_0(x)=e^{i\delta q_{lin}x}$ and $x_1=x-v_s^0t$.
${\bf A}^0[z;q+\delta q_{lin}]$ is a vector, that can be found from
the matrix given by the boundary conditions \cite{Chaplik}.

Eq.~\ref{App23} contains first and second spatial derivatives
and  can be written as
$\hat{L}_{lin} {\bf A}_{lin}=e^{iqx_1}
[f_0(x)\hat{L}_0(q){\bf A}^0+f_0'(x)\hat{L}_1(q){\bf A}^0+
f_0''(x)\hat{L}_2(q){\bf A}^0]=0$,
where $f'=df/dx$. Neglecting the second derivative
$f_0''(x)$, that is $\sim\delta q_{lin}^2\sim K_{eff}^4$, we have

\begin{eqnarray}
\hat{L}_0(q){\bf A}^0+i\delta q_{lin}\hat{L}_1(q){\bf A}^0=0.
\label{App25}
\end{eqnarray}

Now we turn to the nonlinear theory, where the equation
$\hat{L} {\bf A}=0$  is also valid. The operator $\hat{L}$ is
determined by the equations similar to Eqs.~\ref{App21}-\ref{App23}
but with  nonlinear
quantity $\Pi_n(x)$, that is defined by $\Pi_n(x)=-en_n(x)/\phi_n(x,d)$.
In the nonlinear case the vector
${\bf A}={\bf A}_0(x,z)+\sum_{n=1,2,...}a_n(x){\bf A}_n(x,z)e^{iq_nx_1}+c.c.$.
Each term in the equation $\hat{L} {\bf A}=0$  should be zero, and so
$\hat{L}a_n(x) {\bf A}_n(x,z) e^{iq_nx_1}\simeq
e^{iq_nx_1}[a_n(x)\hat{L}_0(q_n){\bf A}_n+ a_n'(x) \hat{L}_1(q_n){\bf A}_n]=0$.
To get the latter equation, we have neglected
$d\delta q_n/dx\sim K_{eff}^4$ and
$d^2a_n/dx^2\sim K_{eff}^4$.
We can solve the equation $\hat{L}{\bf A}=0$ if we choose
${\bf A}_n(x,z)={\bf A}^0[z,q_n+\delta q_n(x)]$,
where the vector ${\bf A}^0[z;q]$ is defined above in the linear theory.
Using Eq.~\ref{App25} we get

\begin{eqnarray}
\frac{da_n(x)}{dx}=i\delta q_n(x)a_n(x),
\label{App26}
\end{eqnarray}
where $\delta q_n(x)$ is given by the equation for $\delta q_{lin}$ (Eq.~\ref{App24})
with  corrections $\Pi_q\rightarrow\Pi_n(x)$ and
$\Pi_q^0\rightarrow\Pi^0_{q_n}=\Pi_n^0$. Eq.~\ref{App26} is used in
Secs.~1,2 to describe the acousto-electric phenomena in a 2DEG.
Using Eqs.~\ref{self} and \ref{j} and the results of Appendix~1,
the denominator in $\delta q_n(x)$ (see Eq.~\ref{App24}) is rewritten as
$1+\Pi_n(x)/\Pi_n^0=-\Pi_n(x)\phi^{saw}_n(x)/ en_n(x)$.
Then, by using Eq.~\ref{App24} and the conservation-of-charge
equation  $j_n(x)=v_s^0en_n(x)$, we obtain

\begin{eqnarray}
\label{dq2}
\delta q_n(x) =-\frac{K^2_{eff} (q_n)}{2}
\frac{2\pi}{\epsilon_{eff}(q_n)} \frac{en_n(x)}{\phi_{n}^{saw}(x)}=
i\frac{K^2_{eff} (q_n)}{2\Pi_n^0v_s^0}
\frac{j_n(x)E_n^{saw*}}{|\phi_{n}^{saw}(x)|^2}.
\end{eqnarray}
We use this equation in Sec.~2 (see Eq.~\ref{dq}).
To obtain above results, we have neglected the terms like
$a_n''(x)$ and $\delta q_n'(x)$ assuming that $K^2_{eff}$
is a small parameter.

\newpage

\section*{Figure captions } 

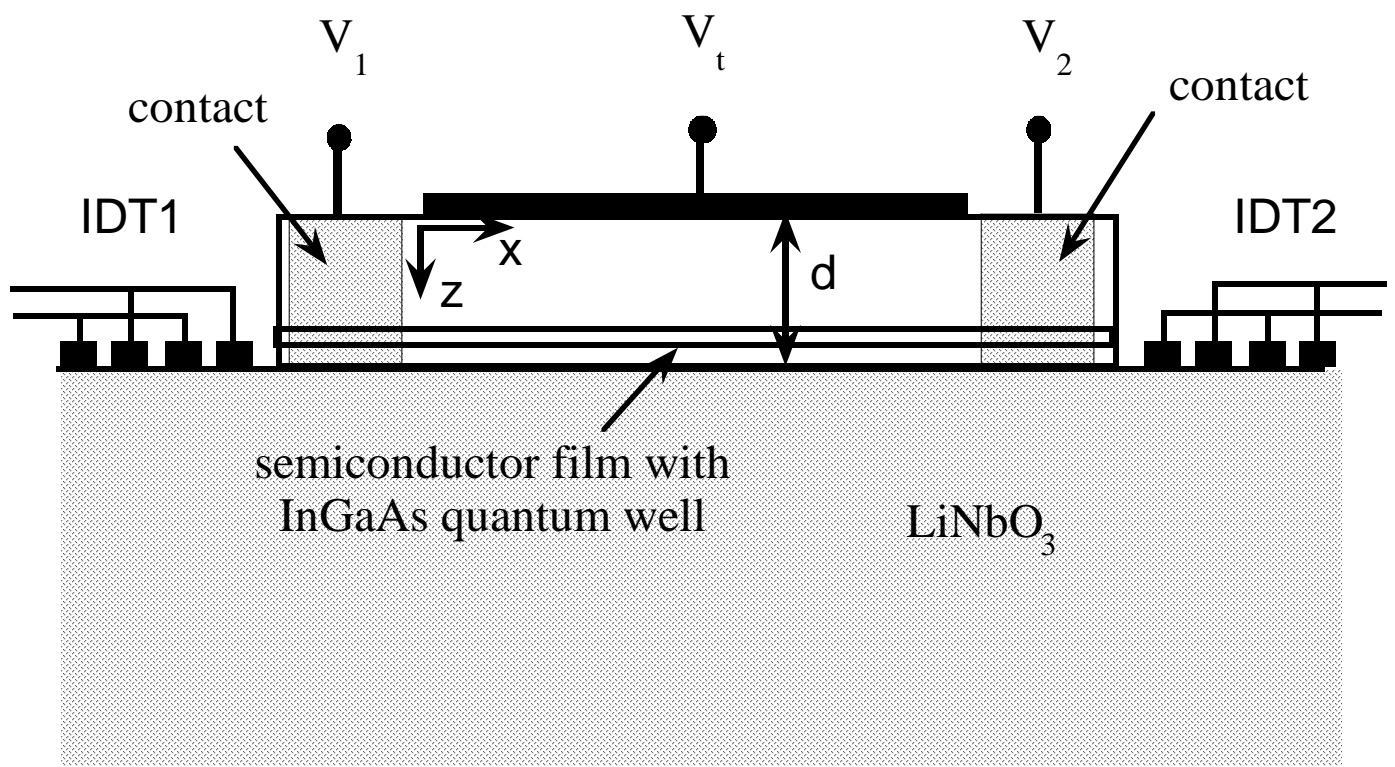
{\bf Fig. 1}. The cross section of a hybrid semiconductor-piezocrystal structure.
An epitaxial lift-off film has a thickness $0.5~\mu m$. The Ohmic contacts
are formed to a 2D electron gas. The transport gate with applied voltage $V_t$ is used to control
the conductivity of the electron plasma. A high-frequency
(RF) signal is applied to the metal interdigital transducer $IDT1$
in order to generate surface acoustic waves. A surface acoustic wave propagates though a
sample
and is detected by the transducer $IDT2$.
\\ \ \\
{\bf Fig. 2}.  The calculated  absorption coefficient
of the first harmonic $\Gamma_1$
as a function of the carrier density $N_s$ for
various potential amplitudes $\Phi_1^{saw}$.
$\lambda=33~\mu m$, $\mu=5000~cm^2V/s$, and $T=300~K$.
Insert: The measured attenuation  of a
SAW as a function of the gate voltage for different
high-frequency (RF) powers applied to the IDT1; $f=114~MHz$.
\\ \ \\
{\bf Fig. 3}. The calculated  absorption coefficient
of the first harmonic $\Gamma_1$ as a function
of the potential amplitude $\Phi_1^{saw}$
induced by a SAW for various fixed densities $N_s$.
The parameters are similar to those in Fig.~2. The dots
show the experimentally measured absorption coefficient
$\Gamma_1$ at the gate voltage $-7.5~V$. This voltage
corresponds to the
maximal attenuation at the smallest RF power.
In the inset we plot the calculated local carrier concentration $n$ as a function of the
in-plane coordinate $x_1$ for different total carrier concentration $N_s$.  The numbers
attached to the plots correspond to $N_s$ in units of $10^{10}~cm^{-2}$.
\\ \ \\
{\bf Fig. 4}.  The calculated SAW-velocity change $\delta v_1$
as a function of the electron density for various potential amplitudes $\Phi_1^{saw}$.
The parameters are similar to those in Fig.~2.
Insert: The measured velocity change of a
SAW as a function of the gate voltage for different
high-frequency (RF) powers applied to the IDT1; $f=114~MHz$.
\\ \ \\
{\bf Fig. 5}.  The calculated intensities of higher harmonics
with $n=2,3$ and $4$ as functions of the
potential amplitude $\Phi_1^{saw}$ for two electron densities
$N_s=10^{10}\ cm^{-2}$ (upper part) and $N_s=10^{11}\ cm^{-2}$
(lower part).  The parameters are similar to those in Fig.~2.
\\ \ \\
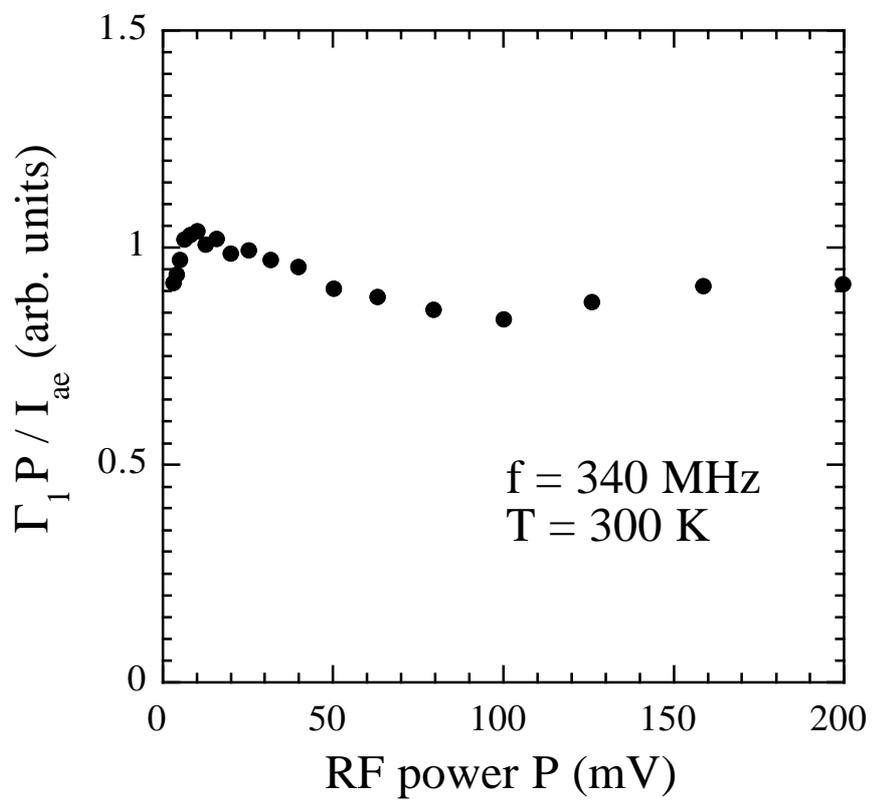
{\bf Fig. 6}.
The measured ratio $\Gamma_1P/I_{ae}$  as a
function of the high-frequency (RF) power $P$ for $f=340~MHz$;
$T=300~K$.  The acousto-electric current  $I_{ae}$
was measured in its maximum.


\begin{thebibliography}{31}
\bibitem{Keff} This formula can be easily obtained in a linear theory
by comparing a general expression for the SAW absorption $<jE>$ and the formula
(1); here $<...>$ means averaging over $\lambda$.
\end{thebibliography}
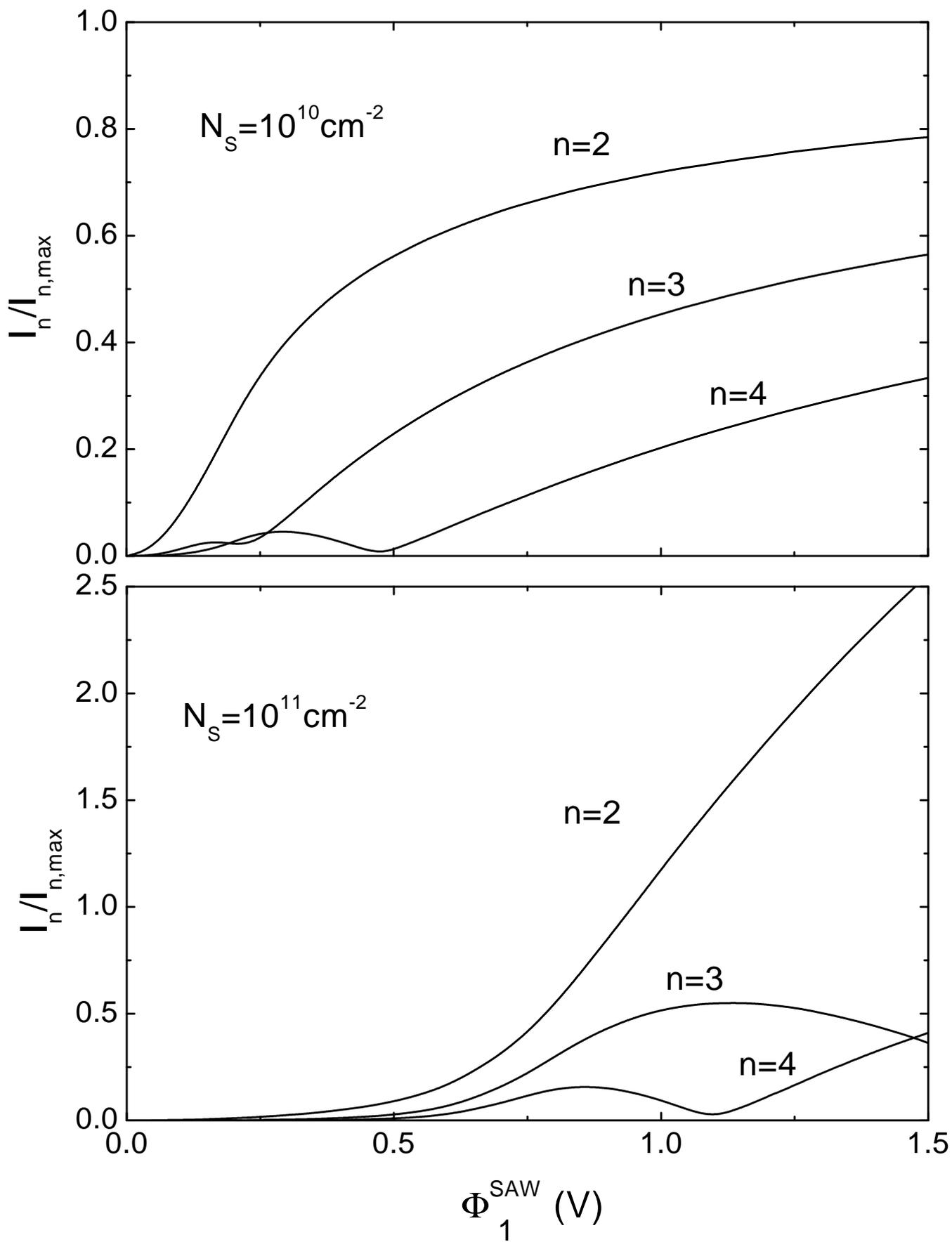
\end{document}